\documentclass[letterpaper, 10 pt, conference]{ieeeconf}  

\IEEEoverridecommandlockouts                              

\usepackage{mathptmx} 
\usepackage{times} 
\usepackage{amssymb}  
\usepackage{graphicx}
\usepackage{amsmath, nccmath}
\usepackage{tabularx,ragged2e,booktabs,caption}
\usepackage{geometry}

\title{\LARGE \bf
Urban Traffic Network Control in Smart Cities; a Distributed Model-based Control Approach}

\author{Roja Eini$^{1}$ and Sherif Abdelwahed$^{2}$
    \thanks{$^{1}$Roja Eini is PhD student of Department of Electrical Engineering,
        Virginia Commonwealth University, Richmond, VA, USA
        {\tt\small einir@vcu.edu}}%
    \thanks{$^{2}$Sherif Abdelwahed is faculty of Department of Electrical Engineering,
        Virginia Commonwealth University, Richmond, VA, USA
        {\tt\small sabdelwahed@vcu.edu}}%
}

\begin{document}

\maketitle
\thispagestyle{empty}
\pagestyle{empty}

\begin{abstract}
This paper proposes a distributed model predictive control (DMPC) approach for an urban traffic network (UTN) system. The control objective is to minimize the traffic congestion and the total travel time spent (TTS) in each link. The proposed DMPC algorithm considers traffic demand and disturbance predictions. The CasADi optimization tool is used to solve the constrained optimization problem. The proposed distributed control approach achieved 60\% less computation time, 14.3\% less TTS, and 15.1\% less queue length compared to the centralized approach. Moreover, while the centralized algorithm neglected the input and state constraints, the distributed approach resulted in the satisfaction of all the constraints over the whole horizon. \\ 

\end{abstract}  

\section{INTRODUCTION}
Traffic congestion is a critical problem with multiple effects on humans’ lives, environment, and society. Several traffic control approaches have been used recently to address the traffic management and optimization problem. 
Model predictive control (MPC) approaches have been widely employed to manage urban traffic networks. The success of MPC in intelligent transport systems, which is a key component in a smart city, arises from its ability to take the predictions about the system’s environmental conditions into account. UTN systems are usually subject to various changes in their structure such as the vehicles’ flow, turning rate, timing, incidents, climatic conditions, and road structure conditions. Recently, several studies have been conducted on using model-based predictive control on UTN systems. However, most of the MPC research on traffic networks are based on the centralized control structure which is impractical in real UTN applications [1-8]. Since UTN systems are associated with large-scale optimization problems, using a centralized strategy demands an extensive computation overhead as well as complex communications. Real-life UTN networks are innately networked systems as they are composed of several lanes with different structures, ramps, slopes, and capacities. Therefore, applying distributed control methodologies to UTN networks is considered a more suitable approach. \\ \indent
In a distributed control approach, the UTN network is decomposed into several subsystems with their local controllers and local objectives. The subsystems interact with each other, and the local controllers exchange their variables through the whole network. Using a distributed control reduces the system’s computational demand and eliminates the need for the system’s global information. In addition, distributed controls are usually more accurate and tolerant to model inaccuracies and system failures.\\ \indent
To the best of our knowledge, a limited amount of literature has addressed the UTN system control problem using distributed MPC approach [9-13]. In [9], a distributed two-layer MPC is developed on a linear traffic model. A distributed MPC based control strategy is proposed in [10] such that the travel time and traffic congestion are reduced. In [11], a distributed MPC approach is implemented on a linear traffic model without taking the system's constraints into account. A hierarchical MPC is developed for a UTN system in [12]. In [13], a distributed MPC scheme is proposed to control a mutil-vehicle platoon. The noted literature applied distributed MPC method either on a nonlinear UTN system without constraints or on a linear UTN system, and some literature (e.g. [12]) did not include the controllers' coordination in the distributed algorithm. The work in [13] also did not consider the external disturbances in the traffic network. Furthermore, none of the related studies considered the traffic inflow demand and the disturbance predictions in the control algorithm.   \\ \indent
This paper introduces a distributed control approach for a nonlinear urban traffic network. The proposed approach is demonstrated on a system with eight junctions and 17 links. The aim is to reduce traffic congestion and total travel time by controlling the signal split time. The main innovation in the proposed approach is utilizing the disturbance and traffic demand predictions. A space-time autoregressive integrated moving average (STARIMA) prediction model is used in the prediction process, and the coordination mechanism is based on the duality Lagrangian functions. The CasADi optimization tool is used to solve the optimization problem, considering the system constraints. The results of the proposed distributed MPC strategy is compared with the results of a centralized MPC on the model. The proposed DMPC strategy improved the system performance significantly. Moreover, the proposed distributed MPC scheme can be implemented in any urban traffic network of this kind.  
\\ \indent
The rest of the paper is organized as follows. Section II describes the UTN model. Section III introduces the centralized and distributed model predictive control approaches. The next section presents the simulation results. Finally, section V provides the conclusions and discusses future research.   
\section{SYSTEM DEFINITION}
\begin{figure}[t]
    \begin{center}
        \framebox{\parbox{3in}{
                \centering
                \includegraphics[height=4cm, width=7cm]{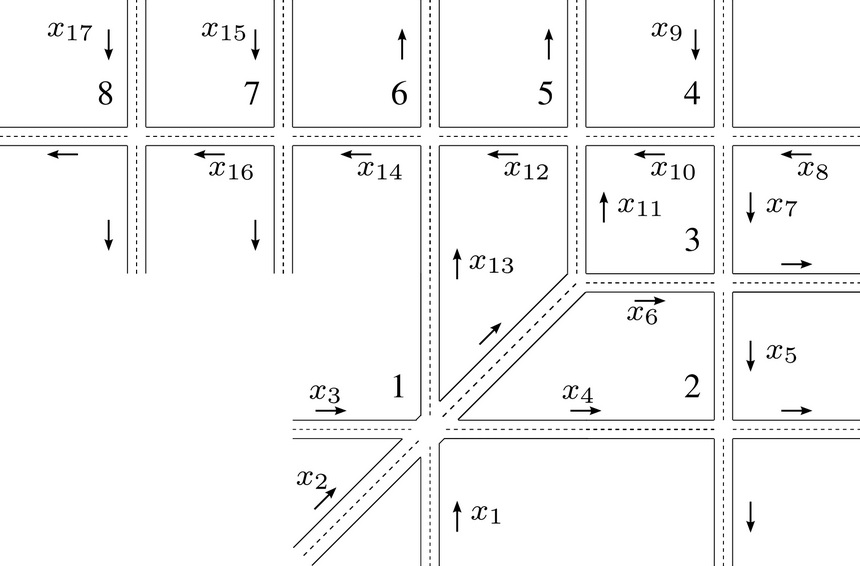}
        }}
        \caption{Urban traffic network configuration}
        \label{fig:UTN_fig}
    \end{center}
\end{figure}

\begin{figure}[t]
    \begin{center}
        \framebox{\parbox{3in}{
                \centering
                \includegraphics[height=4cm, width=7cm]{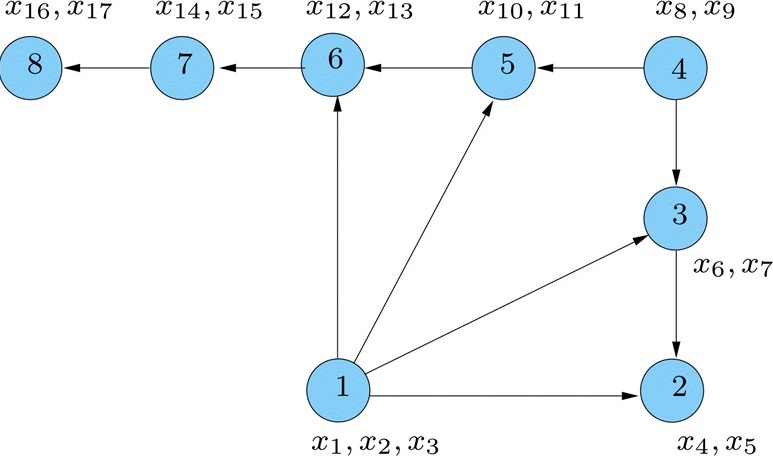}
        }}
        \caption{Urban traffic network graph}
        \label{fig:UTN_graph}
    \end{center}
\end{figure}

A benchmark problem of the nonlinear urban traffic network (Fig. \ref{fig:UTN_fig}) is considered in this work [3]. The UTN is composed of 8 intersections and 17 links. The road links, the intersections, and their associated states are numbered. Inputs are the links’ green time, and the states are the number of vehicles in the links. Fig. \ref{fig:UTN_graph} illustrates the coupling graph of the network. 
In this graph, we assume link z is connecting junctions $j$ and $i$, and there are sets of upstream links $u$ and downstream links $d$ associated with link $z$.

 \indent

\begin{table}
    \begin{center}
        \caption{System Parameters Description}
        \begin{tabular}{| c | m{6cm}|}
            \hline
            $\mathbb{J}$ & {Set of all the junctions in the road} \\  \hline
            $\mathbb{E}$ & {Set of all the roads/links in UTN system} \\ \hline
            $\mathbb{M}$ & {Set of all the subsystems} \\ \hline
            $k$ & {Simulation step} \\ \hline
            $D(z) \in \mathbb{E}$ & {Set of the downstream  junctions of link $z$} \\ \hline
            $U(z) \in \mathbb{E}$ & {Set of the upstream junctions of link $z$} \\ \hline
            $N(m) \in \mathbb{M}$ & {Set of the neighbor subsystems of subsystem $m$} \\ \hline
            $x_z(k)$, $z \in \mathbb{E}$ & {State of link $z$; the number of vehicles in link $z$ at step $k$} \\ \hline
            $q_z(k)$, $z \in \mathbb{E}$ & {Queue length in link $z$ at step $k$} \\ \hline
            $q_{z,d}(k)$, $z, \,\, d \in \mathbb{E}$ & {Queue length of the sub-stream in link $z$ turning to $d$ at step $k$} \\ \hline
            $\tau_d$ & {Sampling time interval}   \\ \hline
            $n_{u,d}(k)$ & {Number of vehicles in link $u$, $d$ at time $k$} \\ \hline
            $S_z(k)$ & {Available storage space of link $z$ at step $k$}  \\ \hline
            ${f_{z}}^{out}(k)$ & {Average flow rate in leaving link $z$ at step $k$}   \\ \hline
            ${f_{z,d}}^{out}(k)$ & {Average flow rate in leaving link $z$ to $d$ at step $k$}   \\ \hline
            ${f_{u,z}}^{out}(k)$ & {Average flow rate in leaving link $u$ to $z$ at step $k$}   \\ \hline
            ${f_{z}}^{arrive}(k)$ & {Average flow rate in reaching the tail of the queue in link $z$ at step $k$}   \\ \hline
            ${f_{z,d}}^{arrive}(k)$ & {Average flow rate in reaching the tail of the queue of sub-strem turning to $d$ in link $z$ at step $k$}   \\ \hline
            ${f_{z}}^{in}(k)$ & {Average flow rate in entering link $z$ at step $k$}   \\ \hline
            ${\mu _{z}}$ & {Saturated flow rate leaving link $z$}   \\ \hline
            $\beta _{z,d}(k)$ & {Relative fraction of the vehicles in link $z$ turning to $d$ at step $k$}   \\ \hline
            ${N_{z}}^{lane}$ & {Number of lanes in link $z$} \\ \hline
            $C_z$ & {Capacity of link $z$}  \\ \hline
            $C_{u,d}$ & {Capacity of link $(u,d)$}  \\ \hline
            ${v_{z}}^{free}$ & {Free-flow vehicle speed in link $z$} \\ \hline
            ${l_{veh}}$ & {Average vehicle length} \\ \hline
            $u_{z,d}(k) \in \mathbb{E}$ & {The green time during step $k$ for the stream in link $z$ toward $d$} \\ \hline
            $c_j$ & {The cycle time of junction $j$} \\ \hline
            $\Delta c_{j,i}$ & {Offset between the cycle time of junction $j$ and $i$} \\ \hline
        \end{tabular}
        \label{tab:systemparams}
    \end{center}
\end{table} 
The state space model of the network is defined as (\ref{eq:UTN model}). \\
\begin{fleqn}
    \begin{equation} 
    \begin{aligned}[b] 
    x_z(k+1)&=x_z(k)+[{f_{z}}^{in}(k)-{f_{z}}^{out}(k)+e(k)]*c_{j}
    \end{aligned}
    \label{eq:UTN model}
    \end{equation}
\end{fleqn} \\ \noindent
where $e(k)$ is the disturbance of link z, i.e. the number of vehicles entering and exiting link $z$. The system parameters are defined in Table \ref{tab:systemparams}. Hence, the inflow rate and outflow rate of link $z$ are defined as (2) and (3), respectively.\\
\begin{fleqn}
    \begin{equation} 
    \begin{aligned}[b] 
    {f_{z}}^{in}(k)&=\sum\limits_{d \in D_z}{} {{f_{z,d}}^{out}(k)}    
    \end{aligned}
    \label{eq:inflow}
    \end{equation}
\end{fleqn} 
\begin{fleqn}
    \begin{equation} 
    \begin{aligned}[b] 
    {f_{z}}^{out}(k)&=\sum\limits_{u \in U(z)}{} {{f_{u,z}}^{out}(k)} 
    \end{aligned}
    \label{eq:outflow}
    \end{equation}
\end{fleqn} \\ \indent
The outflow rate, ${f_{z}}^{out}$, is updated through (\ref{eq:outflow_update}) in each step. \\
\begin{fleqn}
    \begin{equation} 
    \begin{aligned}[b] 
    {f_{z}}^{out}(k+1)&={f_{z}}^{out}(k)+[{f_{z,d}}^{arrive}(k)-{f_{z,d}}^{out}(k)]*c_{j}
    \end{aligned}
    \label{eq:outflow_update}
    \end{equation}
\end{fleqn} \\ \indent
Moreover, the arrival flow rate, ${f_{z,d}}^{arrive}$, is defined as (\ref{eq:arriveflow_z,o}). \\
\begin{fleqn}
    \begin{equation} 
    \begin{aligned}[b] 
    {f_{z,d}}^{arrive}(k)&={f_{z}}^{arrive}(k)*\beta _{z,d}(k)
    \end{aligned}
    \label{eq:arriveflow_z,o}
    \end{equation}
\end{fleqn} \\
\noindent
where the delayed flow rate, ${f_{z}}^{arrive}$, arriving the tail of the queue is attained through (\ref{eq:arriveflow_z}).\\
\begin{fleqn}
    \begin{equation} 
    \begin{aligned}[b] 
    {f_{z}}^{arrive}(k)&=(\frac {c_{j}- \gamma (k)}{c_j})* {f_{z}}^{in}(k-\delta (k))  \\
    &+\frac{\gamma(k)}{c_j}* {f_{z}}^{in} (k- \delta(k)-1)
    \end{aligned}
    \label{eq:arriveflow_z}
    \end{equation}
\end{fleqn} \\
\noindent 
where $\delta (k)$ and $\gamma (k)$ are attained from the time delay equations in (\ref{eq:time_delay_formulas}). \\
\begin{fleqn}
    \begin{equation} 
    \begin{aligned}[b] 
    \delta (k)&=\lfloor \frac {\tau (k)}{C_z} \rfloor  \\
    \gamma (k)&= rem(\tau (k), \tau _d),\,\,\,\,\,\,  \tau (k)=\frac{(C_z - q_z(k))* l_{veh}}{{N_{z}}^{lane}*{v_{z}}^{free}*c_j}
    \end{aligned}
    \label{eq:time_delay_formulas}
    \end{equation}
\end{fleqn} \\ \indent
The number of vehicles waiting in queue for turning to downstream links $d$ is updated through (\ref{eq:queue_update}). \\
\begin{fleqn}
    \begin{equation} 
    \begin{aligned}[b] 
    q_{z,d}(k+1)&=q_{z,d}(k)+[{f_{z,d}}^{arrive}(k)-{f_{z,d}}^{out}(k)]*c_{j}
    \end{aligned}
    \label{eq:queue_update}
    \end{equation}
\end{fleqn} \\ \indent
Having (\ref{eq:arriveflow_z,o}) and (\ref{eq:queue_update}), (\ref{eq:inflow}) can be converted to (\ref{eq:inflow_minimum}). Three different scenarios are considered in (\ref{eq:inflow_minimum}): unsaturated, saturated, and over-saturated traffic flows.  \\
\begin{fleqn}
    \begin{equation} 
    \begin{aligned}[b] 
    {f_{z}}^{in}(k)&=min\{ (\frac{q_{z,d}(k)}{c_j} + {f_{z,d}}^{arrive}(k)), (\frac{\beta _{z,d}(k)*{\mu _{z}}*u_{z,d}(k) }{c_j}),\\ &((\frac{\beta _{z,d}(k)}{\sum\limits_{u \in U(z)}{} \beta _{z,d}(k)}) *(\frac{C_{u,d}-n_{u,d}(k)}{c_j}))                            \}
    \end{aligned}
    \label{eq:inflow_minimum}
    \end{equation}
\end{fleqn} 
\indent
Note that the cycle times $c_j$ and $c_i$ can be different from each other, and the offset time $\Delta c_{j,i}$ can be defined for junctions $i$ and $j$. So the simulation steps for intersections $j$ and $i$ are denoted as $k_j$ and $k_i$, respectively, which are not necessarily equal. Therefore, to synchronize the inflow and outflow rates the discrete-time outflow rate of upstream links is transformed into its continuous-time form by sampling counts of $k_i$. Then, it is converted to its discrete-time form by sampling counts of $k_j$ as (\ref{eq:intersection sampling_synchronization}).  \\
\begin{fleqn}
    \begin{equation} 
    \begin{aligned}[b] 
    {f_{u,z}}^{out}(t)&= {f_{u,z}}^{out}(k_j), \,\,\,\,\,\, k_j * c_j  \leq  t < (k_j +1) * c_j \\
    {f_{u,z}}^{out}(k_i)&= \frac{}{} \int\limits_{k_m*c_m+\Delta c_{j,i}}^{(k_j +1)*c_j+\Delta c_{j,i}}{{{f_{u,z}}^{out}(t)dt}}    
    \end{aligned}
    \label{eq:intersection sampling_synchronization}
    \end{equation}
\end{fleqn} 
\indent 
It is required to define a common control time interval for all the subnetworks so that they can communicate synchronously. $T_c$ is defined as an integer $N$ times the least common multiple, $T_{lcm}$, of all the intersections' time intervals. \\
\begin{fleqn}
    \begin{equation} 
    \begin{aligned}[b] 
    T_c&=N*T_{lcm}\\
    k_c(k_j)&=\lfloor \frac{k_j}{N*N_j}  \rfloor,\,\,\,\,\,\,j \in \mathbb{J}, \,\,\,\,\,\, N_j \,\,\text{and}\,\, N \,\,\text{are integers}
    \end{aligned}
    \label{eq:control_time_interval}
    \end{equation}
\end{fleqn} \\
\section{THE MODEL PREDICTIVE CONTROL APPROACH}
This section introduces the control approach for the traffic system discussed in the previous section. We first introduce the centralized approach and then extend this structure to a distributed control in which a set of local controllers are coordinating toward a given global objective. 
\subsection{Centralized MPC}
\begin{figure}[thpb]
    \begin{center}
        \framebox{\parbox{3in}{
                \centering
                \includegraphics[height=5cm, width=7cm]{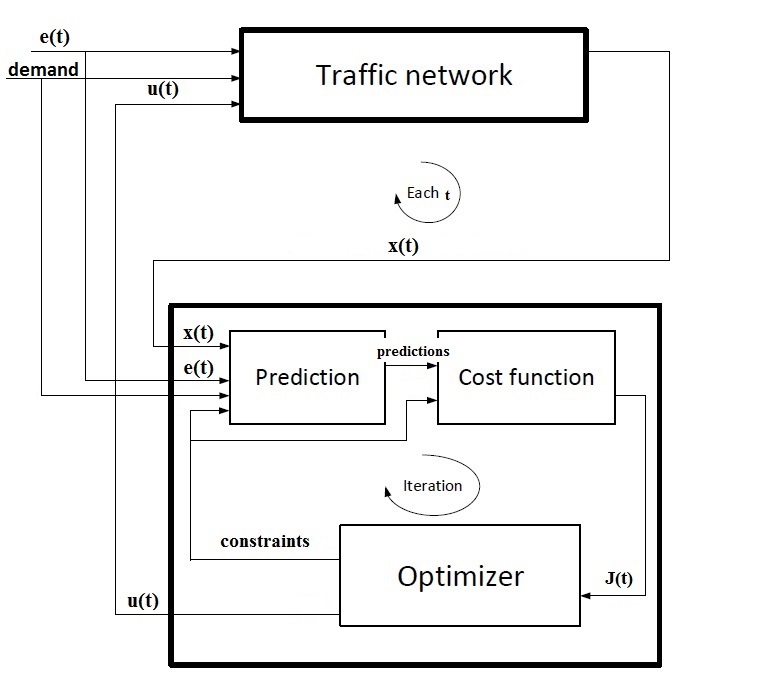}
        }}
        \caption{Centralized MPC and the UTN system block diagram}
        \label{fig:MPC}
    \end{center}
\end{figure}

The centralized cost function is defined as (\ref{eq:centlazied_cost}), containing the states (number of vehicles in links) and control signal (total green time). \\
\begin{fleqn}
    \begin{equation} 
    \begin{aligned}[b] 
    J(k)&=min \{\sum\limits_{p=1}^{K_p} \sum\limits_{z \in \mathbb{E}} {\lVert{\hat{x}_z(k_j+p|k)\lVert}_Q }^2 \\
    &+ \sum\limits_{p=1}^{K_c} {\sum\limits_{z \in \mathbb{E}}} {\lVert{\hat{u}_z(k_u+p|k)\lVert}_R }^2 \}
    \end{aligned}
    \label{eq:centlazied_cost}
    \end{equation}
\end{fleqn} \\ \indent
$Q$ and $R$ are the weighting matrices associated with the system states and the control signals, respectively [14], [15]. $\hat{u}$ and $\hat{x}$ are the predicted control signal and the predicted state up to the control horizon $K_c$ and prediction horizon $K_p$, respectively. The constraints are the model constraints (\ref{eq:centlazied_prediction_system_model}), the inflow traffic demand constraints (\ref{eq:demand_constraints}), and the control signal constraints (\ref{u costraints_centralized}). Note that the input constraints implicitly include the disturbance ($e_z$) constraints.\\
\begin{fleqn}
    \begin{equation} 
    \begin{aligned}[b] 
    &\hat{x}_z(k_j+p+1) = f(\hat{x}_z(k_j+p),u_z(k_c+p),d_z(k_j+p))  \\
    &\hat{x}_z(k_j) = {x}_z(k_j)   
    \end{aligned}
    \label{eq:centlazied_prediction_system_model}
    \end{equation}
\end{fleqn} 
\begin{fleqn}
    \begin{equation} 
    \begin{aligned}[b] 
    \hat{f}_{u,z}^{out}(k_j) &= {f}_{u,z}^{out}(k_j) 
    \end{aligned}
    \label{eq:demand_constraints}
    \end{equation}
\end{fleqn} 
\begin{fleqn}
    \begin{equation} 
    \begin{aligned}[b] 
    & \sum\limits_{z \in IN(j)}{u_z} + L_z = C  \\
    & {u_z}^{min} \leq u_z \leq {u_z}^{max} 
    \end{aligned}
    \label{u costraints_centralized}
    \end{equation}
\end{fleqn} \\ \indent
In above, $\hat{f}_{u,z}^{out}$, $C$, and $L_z$ denote the predicted traffic inflow demand, whole cycle time, and the total lost time of link $z$, respectively. The control signal constraints (the minimum and maximum green time) are to maintain the safety and the total cycle time. Furthermore, the last term of (\ref{eq:inflow_minimum}) associated with over-saturated condition can be omitted by adding the state constraints (\ref{extra_over_saturated_constraints}) to the optimization problem. In result, the number of vehicles in a link would not exceed the link capacity, and over-saturated condition will not happen. \\
\begin{fleqn}
    \begin{equation} 
    \begin{aligned}[b] 
    0 \leq x_z(k)\leq C_z
    \end{aligned}
    \label{extra_over_saturated_constraints}
    \end{equation}
\end{fleqn} 
\indent
The  STARIMA prediction model (\ref{starima}) is applied to predict the disturbances and traffic demand. STARIMA is an extension of autoregressive integrated moving average (ARIMA) model in which the spatial dimension of the UTN is taken into account. It is proved that prediction results will be significantly improved by including the spatial dimension in the prediction model [16]. \\
\begin{fleqn}
    \begin{equation} 
    \begin{aligned}[b] 
    \nabla^{d}\hat{x}(t) &=\sum_{i=1}^{p}\sum_{k=0}^{m_i}\emptyset_{ik}{\rm W}_{k}\nabla^{d}\vec{x}(t-i)-\sum_{i=1}^{q}\sum_{k=0}^{n_i}\theta_{ik}{\rm W}_{k}\vec{e}(t-i)+\vec{e}(t) 
    \end{aligned}
    \label{starima}
    \end{equation}
\end{fleqn} 
\indent
The $N$ by $N$ dimension weight matrix $W_k$ indicates spatial dimension factor of the network. $\phi _ {ik}$ and $\theta _{ik}$ denote the auto-regression parameter and the moving average parameter, respectively. $p$ and $q$ are the time lags, and $d$ is the degree of differencing. $\hat{x}(t)$ is the detected signal at time $t$, $m_i$ is the spatial order of the $i$th autoregressive term, and $n_i$ is the spatial order of the moving average term. \\ \indent
Hence, the cost function in the centralized framework is (\ref{eq:centlazied_cost}) subject to the model constraints (\ref{eq:centlazied_prediction_system_model}), demand constraints (\ref{eq:demand_constraints}), input constraints (\ref{u costraints_centralized}), and state constraints (\ref{extra_over_saturated_constraints}). Fig. \ref{fig:MPC} shows the block diagram of the system and the centralized MPC controller. The centralized MPC algorithm for UTN system is as follows. \\

\begin{itemize}
    \item At step $k=0$, determine the state value $x(0)$, and get the input value $u(0)$ by solving the optimization problem. 
    \item At step $k>0$, get the state, input, traffic demand, and disturbance predictions from (\ref{starima}) (with $K_c$ as the disturbance and input horizon, and $K_p$ as the state horizon).
    \item At step $k>0$, solve the optimization problem (\ref{eq:centlazied_cost}) to get the optimum input signal through the control horizon.
    \item Apply the optimum control signal at step $k$ to the system, and get the updated state values.
    \item $k=k+1$, and go to the second step.
\end{itemize} 

\indent
The disturbance, traffic demand, output, and input predictions are attained in each step of the algorithm. However, the centralized algorithm requires excessive computation since, in each step, a large optimization problem with many variables is being solved.
\subsection{Distributed MPC}
\begin{figure}[thpb]
    \begin{center}
        \framebox{\parbox{3in}{
                \centering
                \includegraphics[height=5cm, width=7cm]{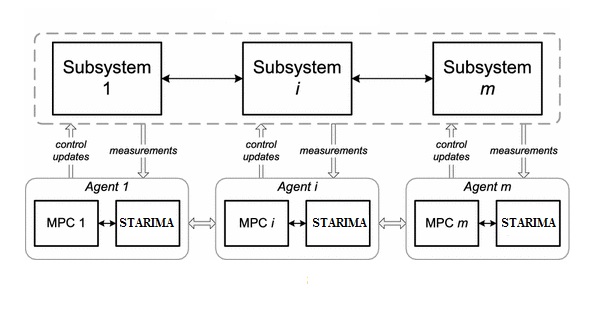}
        }}
        \caption{Distributed MPC of m interacted subsystems with controllers' coordination}
        \label{MPC_dis}
    \end{center}
\end{figure}

In the proposed distributed approach, one local MPC controller is assigned to each subsystem. The goal is to achieve a specific global system objective. Fig. \ref{MPC_dis} shows a diagram of a distributed MPC for a system with m subsystems [17], [18].
 
\indent
To apply distributed MPC on the traffic network, the system is decomposed into $M$ subsystems. $m$ is regarded as a subsystem if and only if it contains a center junction $j$ with connecting links. Hence, the dynamic equation of subsystem $m$ is stated as (\ref{eq:distributed_system_model}). Note that only the neighboring subsystems interactions are considered in each subsystem’s model. In (\ref{eq:distributed_system_model}), $i,m \in \mathbb{M}$ are the neighboring subsystems with interaction equations $D_{im}$. \\
\begin{fleqn}
    \begin{equation} 
    \begin{aligned}[b] 
    x_m(k+1)&=x_m(k)+\sum\limits_{i \in N(j)} D_{im}(k) - f_{out,m}(x_m(k),u_m(k))
    \end{aligned}
    \label{eq:distributed_system_model}
    \end{equation}
\end{fleqn} 
\indent
Above, $x_m$ and $u_m$ are the state and input vectors of subsystem $m$. $f_{out,m}$ is the outflow of subsystem $m$. Moreover, the cost function is decomposed into sub cost functions $\phi_m$ for each subsystem $m$. Therefore, the general cost function and each sub cost function is stated as (\ref{eq:distributed_cost}). \\
\begin{fleqn}
    \begin{equation} 
    \begin{aligned}[b] 
    & J(k) = min \{\sum\limits_{m \in \mathbb{M}} \phi_m \}\\
    & {\phi_m}(k) = \{\sum\limits_{p=1}^{K_p}  {\lVert{\hat{x}_m(k+p|k)\lVert}_{Q_m} }^2 + \sum\limits_{p=1}^{K_c}  {\lVert{\hat{u}_m(k+p|k)\lVert}_{R_m}^2} \}
    \end{aligned}
    \label{eq:distributed_cost}
    \end{equation}
\end{fleqn} 
\indent
In (\ref{eq:distributed_cost}), $Q_m$ and $R_m$ are the weighting matrices associated with the state and input vectors of subsystem $m$, respectively. Also, the distributed problem constraints are the model constraints (\ref{eq:centlazied_prediction_system_model}), the demand constraints (\ref{eq:demand_constraints}), the input constraints (\ref{u costraints_centralized}), and the state constraints (\ref{extra_over_saturated_constraints}). The control objective is to minimize the cost function containing the system's inputs and states.\\ \indent
The coordination mechanism of the proposed distributed MPC approach is based on the dual decomposition method [19]. The Lagrangian multipliers are used to impose the system model, the input signal and the state constraints into the optimization cost function. In result, there is only one optimization problem without constraints, and the solution to this problem is approximately equal to the solution to the original problem subject to constraints. Thus, the augmented function for subsystem m is stated as (\ref{augmented_cost}). \\
\begin{fleqn}
    \begin{equation} 
    \begin{aligned}[b] 
    {L}_{m}(k)&= {\phi_m}(k) + \sum\limits_{{i,m}\in N(m)}{} [{\lambda_{im}(k)}^T (D_{im}(k)-\hat{D}_{im}(k)) \\
    &+ \frac{\rho}{2}  {\lVert D_{im}(k)-\hat{D}_{im}(k) \lVert_{2}^2}]  \\
    \end{aligned}
    \label{augmented_cost}
    \end{equation}
\end{fleqn} 
where $\hat{D}_{im}$ is the predicted interaction between neighboring subsystems $i$ and $m$, and $\lambda_{im}$ is the duality function coefficient.  \\
\indent
At each time step, each subsystem only communicates with its neighbors in $N(m)$ to get the interconnecting information from them, and send its last updated state and input variables to them. Therefore, the optimization problem for each local controller would be the minimization of (\ref{augmented_cost}), subject to constraints (\ref{u costraints_centralized}). It is worth mentioning that the duality function coefficients $\lambda_{im}$ should be optimized as well as the input signal in each iteration, through (\ref{lambda}). \\
\begin{fleqn}
    \begin{equation} 
    \begin{aligned}[b] 
    \lambda_{im}^{s+1}(k) &= \lambda_{im}^{s}(k)+ \alpha ^{s} (D_{im}^s(k)-{\hat{D}_{ij}^s(k)})
    \end{aligned}
    \label{lambda}
    \end{equation}
\end{fleqn} 
\indent
The local cost functions are modified in the coordination procedure such that they include the global cost function of the whole system. In fact, the impact of input signal $u_m$ on the global objective function $J(k)$ is considered through the interaction term $D_{im}$. By defining $\mu_{mi}$ parameter as (\ref{mu}), the global cost function changes can be added to the local cost of each subsystem as (\ref{updated_distributed_cost}). \\
\begin{fleqn}
    \begin{equation} 
    \begin{aligned}[b] 
    &\mu_{mi}(k) = \frac{d\phi_i(t)}{du_m} \\
    &\phi_i(k) = \{\sum\limits_{p=1}^{K_p}  {\lVert{\hat{x}_i(k_j+p|k_j)\lVert}_{Q_i} }^2 + \sum\limits_{p=1}^{K_c}  {\lVert{\hat{u}_i(k_c+p|k_c)\lVert}_{R_i}^2} \} 
    \end{aligned}
    \label{mu}
    \end{equation}
\end{fleqn} 
\begin{fleqn}
    \begin{equation} 
    \begin{aligned}[b] 
    &{\phi_m}^{updated}(k) = \phi_m(k) + \sum\limits_{i \in \mathbb{M} \,\, i\neq m} {\mu_{mi}(k)(u_m - {u_m}^{opt})}
    \end{aligned}
    \label{updated_distributed_cost}
    \end{equation}
\end{fleqn} 
\indent
Thus, the following distributed MPC algorithm is proposed for the nonlinear traffic network problem.\\
Step 1: 
\begin{itemize} \nonumber
    \item Send $u_m(k-1|k-1)$ and $\hat{x}_m(k|k-1)$ to its neighboring controller $C_i$ (coordination mechanism). 
    \item Estimate the future state trajectories $\hat{x}_i(k|k-1)\,\,\, i\neq m $ and control inputs $u_i(k-1|k-1)\,\,\, i\neq m$ from its neighboring controller through information exchange (\ref{eq:distributed_system_model}).
    \item Observe the values of $x_m(k)$.
    \item Build $\hat{x}(k|k-1)$ and $u(k|k)$ by adding the subsystem's state estimations $\hat{x}_m(k|k-1)$, control inputs $u_m(k|k)$, and subsystem's neighbor information $\hat{x}_i(k|k-1)$, and $u_i(k-1|k-1)$ to attain the predictions of $\hat{D_{im}}(k|k-1)$.
\end{itemize}
Step 2:  
\begin{itemize} 
    \item Calculate the optimal control law $u_m(k|k)$ by solving the optimization problem (\ref{augmented_cost}) through CasADi optimization tool.
    \item Apply the first element of the optimal control $u_m(k|k)$ to the system, and update the local cost function through (\ref{updated_distributed_cost}).
    \item Update $\lambda$ from (\ref{lambda}).
\end{itemize}
Step 3: 
\begin{itemize} 
    \item Compute the estimation of the future state, disturbance, and traffic demand trajectory of $m$th subsystem over the horizon $K_p$ from the prediction model (\ref{starima}).
\end{itemize}
Step 4:  
\begin{itemize} 
    \item replace $k$ by $k+1$, and go to step 1 to repeat the algorithm.
\end{itemize} 

\indent
The coordination strategy in the proposed distributed algorithm based on duality function avoids global communication in the whole network [19]. Moreover, the distributed approach requires less computation, demonstrates improved overall performance, and considers all the constraints.\\
\section{SIMULATION RESULTS}
The control horizon, $K_c$, and prediction horizon, $K_p$, are chosen as 4 and 7, respectively. Moreover, the weight matrices Q and R are chosen as (\ref{QandR}). In the distributed approach, the weighting matrices are defined for each local controller. \\
\begin{fleqn}
    \begin{equation} 
    \begin{aligned}[b] 
    Q&=0.1\times {I_{17K_p\times17K_p}}, \,\,\,\,\,\,\,\,R=0.3\times {I_{17K_c\times17K_c}}
    \end{aligned}
    \label{QandR}
    \end{equation}
\end{fleqn}  \indent
At each time instant, the optimization problem is solved using CasADi tool in MATLAB, then the optimum input is considered as the current input for the next step. Minimizing the total time spent (TTS) criterion is the control objective which corresponds to the accumulated amount of time spent by the vehicles. The simulations are performed for 4 hours, with the sampling rate of 10 seconds. The simulation assumptions are as follows. \\
- The cycle time for all the junctions is 120 s. \\
- The vehicles' speed in the free flow rate is 40 km/h. \\
- The traffic supply flow rate (demand) is a balanced Gaussian distribution function between 300 and 800. \\
- The maximum and minimum values for the green time are 10 and 520 seconds respectively. \\
- The capacity of each link is 1000. \\ \indent
For the centralized MPC case, the cost function is a global function as (\ref{eq:centlazied_cost}), and the control law is calculated at each time step through the centralized algorithm. \\ \indent 
Based on the UTN graph in Fig. \ref{fig:UTN_graph}, 8 subsystems corresponding to 8 intersections are considered in the distributed control case. Moreover, each subsystem has one objective function ${\phi_m}(k)$ as (\ref{eq:distributed_cost}). The interactions of subsystem $m$ with its neighboring subsystem $i$ ($D_{im}(k)$) is stated using the turning rate and the outflows. For instance, subsystem 5 has two interactions $D_{15}(k)$ and $D_{45}(k)$ with subsystem 1 and 4 respectively. Thus, the dynamic interaction equations are stated as (\ref{D15}).  \\
\begin{equation}  \nonumber
\begingroup 
\setlength\arraycolsep{0.5pt} 
\begin{bmatrix} 
x_{10}(k+1) \\
x_{11}(k+1)
\label{x10_x11}
\end{bmatrix} 
=
\begin{bmatrix} 
x_{10}(k) \\
x_{11}(k)
\end{bmatrix}
+D_{15}(k)+D_{45}(k)-f_{out,5}
\endgroup
\end{equation}
\begin{equation}  \nonumber
\begingroup 
\setlength\arraycolsep{0.5pt}
D_{15}(k)= 
\begin{bmatrix} 
\tau_{1,10}\,\,\,q_{out,1} \\
0
\end{bmatrix} 
+
\begin{bmatrix} 
\tau_{2,10}\,\,\,q_{out,2}  \\
0
\end{bmatrix} 
+
\begin{bmatrix} 
\tau_{3,10}\,\,\,q_{out,3}  \\
0
\end{bmatrix} 
\endgroup
\label{D15}
\end{equation}
\begin{equation}  \nonumber
\begingroup 
\setlength\arraycolsep{0.5pt}
D_{45}(k)= 
\begin{bmatrix} 
0 \\
\tau_{1,11}\,\,\,q_{out,1}
\end{bmatrix} 
+
\begin{bmatrix} 
0  \\
\tau_{2,11}\,\,\,q_{out,2}
\end{bmatrix} 
+
\begin{bmatrix} 
0  \\
\tau_{3,11}\,\,\,q_{out,3}
\end{bmatrix} 
\endgroup
\label{D45}
\end{equation}
\begin{equation}  
\begingroup 
\setlength\arraycolsep{0.5pt} 
f_{out,5}=
\begin{bmatrix} 
q_{out,10}(x_{10}(k),u_{10}(k)) \\
q_{out,11}(x_{11}(k),u_{11}(k))
\label{fout5}
\end{bmatrix} 
\endgroup
\end{equation} 
\indent
Without loss of generality, the interaction parameter ($\mu_{mi}$) for subsystem 5 is calculated in (\ref{phi5_mu_3}). $\phi _5$ corresponding to subsystem 5 cost function is composed of the input signal $\hat{u}_{1k},\,\,\, k=1,\cdots,K_c$. The control vector of subsystem 1 (subsystem 5 neighbor) is ${[u1,\,\,\,u2,\,\,\,u3]}^T$.  \\ 
\begin{equation}  \nonumber
\begingroup 
\setlength\arraycolsep{1pt} 
\mu_{5i} = \sum\limits_{s=k}^{K_c}\{{[\frac{d\phi_5}{d\hat{x}_{10}},\,\,\,\frac{d\phi_5}{d\hat{x}_{11}}]}
\begin{bmatrix} 
\frac{d\hat{x}_{10}}{d{D_{15}^1}} & \frac{d\hat{x}_{10}}{d{D_{15}^2}}\\
\frac{d\hat{x}_{11}}{d{D_{15}^1}} & \frac{d\hat{x}_{11}}{d{D_{15}^2}}\\
\end{bmatrix}
\}
\begin{bmatrix} 
\frac{d{D_{15}^1}}{d\hat{u}_1} & \frac{d{D_{15}^1}}{d\hat{u}_2} & \frac{d{D_{15}^1}}{d\hat{u}_3}\\
\frac{d{D_{15}^2}}{d\hat{u}_1} & \frac{d{D_{15}^2}}{d\hat{u}_2} & \frac{d{D_{15}^2}}{d\hat{u}_3}\\
\end{bmatrix}
\label{phi5_mu_1} 
\endgroup
\end{equation} 
\begin{equation}  
\begingroup 
\setlength\arraycolsep{1pt} 
=[\sum\limits_{s=k}^{K_c}{2Q_5\hat{x}_5(t+s)}]^T
\begin{bmatrix} 
\frac{d{D_{15}^1}}{d\hat{u}_1} & \frac{d{D_{15}^1}}{d\hat{u}_2} & \frac{d{D_{15}^1}}{d\hat{u}_3}\\
0 & 0 & 0\\
\end{bmatrix}
\label{phi5_mu_3} 
\endgroup
\end{equation} 
\indent
The derivatives of the interaction equations $D_{im}$ versus the input signals can be behaved as (\ref{D_im_derivatives}). \\
\begin{fleqn}
    \begin{equation} 
    \begin{aligned}[b] 
    \frac{d{D_{15}^1}}{du_{i}} &= 0  \,\,\,\,\,\,\, if \,\,\,\,\, x_i<u_i(\frac{S_i}{C_i})\,\, , \,\,\,\, i=1, \cdots,3 \\
    \frac{d{D_{15}^1}}{du_{i}} &= \tau_{i,10}(\frac{S_i}{C_i}) \,\,\,\,\,\,\, if \,\,\,\,\, x_i>u_i(\frac{S_i}{C_i})\,\, , \,\,\,\, i=1, \cdots,3
    \end{aligned}
    \label{D_im_derivatives}
    \end{equation}
\end{fleqn} 
\indent
Subsystem $m$ cost function is derived from (\ref{eq:distributed_cost}) as (\ref{eq:distributed_cost_m}). \\
\begin{fleqn}
    \begin{equation} 
    \begin{aligned}[b] 
    {\phi_m}(k) &= \{\sum\limits_{p=1}^{K_p}  {\lVert{\hat{x}_m(k+p|k)-{{x}_m}^d(k+p|k)\lVert}_{Q_m} }^2  \\
    &+\sum\limits_{p=1}^{K_c}  {\lVert{\hat{u}_m(k+p|k)\lVert}_{R_m}^2} \} \\
    & {x_m}^d:\,\, \text{desired state trajectory for subsystem $m$}
    \end{aligned}
    \label{eq:distributed_cost_m}
    \end{equation}
\end{fleqn}
\indent

\begin{figure}[thpb]
    \begin{center}
        \framebox{\parbox{3.2in}{
                \centering
                \includegraphics[height=6.5cm, width=8.7cm]{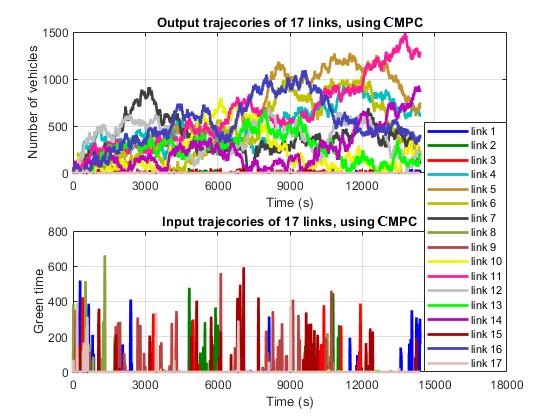}
        }}
        \caption{17 links state and input trajectories using centralized MPC}
        \label{fig:centralized}
    \end{center}
    \begin{center}
        \framebox{\parbox{3.2in}{
                \centering
                \includegraphics[height=6.5cm, width=8.7cm]{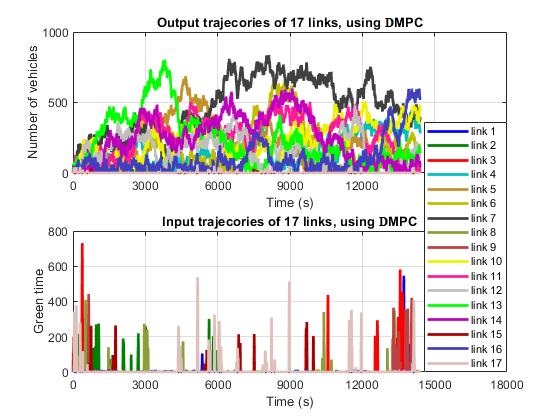}
        }}
        \caption{17 links state and input trajectories using distributed MPC}
        \label{fig:distributed}
    \end{center}
\end{figure}  
\begin {table} [t]
\caption {Simulation results}
\begin{center}
    \noindent
    \newlength{\myheight}
    \setlength{\myheight}{0.5cm}
    \footnotesize
    \begin{tabular}{|l|c|c|c|l|}  
        \hline
        \parbox[c][\myheight][c]{0cm}{}   Parameters \small &  CMPC  \small & DMPC  \small & Change \small \\  
        \hline \hline 
        \parbox[c][\myheight][c]{0cm}{}    Maximum queue length  \small & 437  \small & 371  \small & $\downarrow$ 15.1\%  \small  
        \\   
        \parbox[c][\myheight][c]{0cm}{}    Maximum number of vehicles \small & 1497 \small & 815 \small &  $\downarrow$ 45.5\% \small
        \\  
        \parbox[c][\myheight][c]{0cm}{}    Total Time Spent (s) \small & 4.0966e+03 \small & 3.5091 \small & $\downarrow$ 14.3\% \small
        \\ 
        \parbox[c][\myheight][c]{0cm}{}   Run time (s) \small &  4630 \small & 1852  \small & $\downarrow$ 60.0\% \small \\  
        \hline \hline 
    \end{tabular}
    \label{tab:Simulation results} 
\end{center}
\end {table}  
\normalsize
Having $D_{im}$ and $\phi_m$ values, augmented function $L_m$ is defined as (\ref{augmented_cost}). Based on the proposed distributed algorithm, optimum input is calculated, and applied to the system. The output/input, disturbance, and demand predictions, $\mu_{mi}$, $D_{im}$, and $L_m$ are updated up to the horizons and then, the next optimum input is calculated. \\ \indent
Figs. \ref{fig:centralized} and \ref{fig:distributed} show the simulation results using the centralized and distributed MPC on the traffic network, respectively. Furthermore, Table \ref{tab:Simulation results} compares the numerical values of TTS, maximum number of vehicles, maximum queue length, and computation time using the two controllers. \\ \indent
As shown in state (output) graphs, the distributed MPC shows slightly less number of vehicles in each link compared to the centralized MPC strategy. Moreover, using the centralized approach, the number of vehicles in links 5 and 11 exceeds the state constraint (1000 vehicles), whereas the distributed method resulted the satisfaction of all the constraints throughout the simulation time. From the first two row values of Table \ref{tab:Simulation results}, the maximum queue length and the maximum number of vehicles are reduced by 15.1\% and 45.5\%, respectively, using the distributed MPC. Besides, the TTS value decreased by 14.3\% using the proposed distributed approach (third row of Table \ref{tab:Simulation results}). Also, comparing the control trajectories of Fig. \ref{fig:centralized} and \ref{fig:distributed}, the distributed approach performs slightly better in minimizing the control signal than the centralized method. Worth mentioning that, the distributed structure performance is slightly better compared to the centralized MPC, since it is more accurate and responds faster due to the lower computational demand.  \\ \indent
Furthermore, the maximum optimization time, in a corei7 3.2GHz computer, for distributed MPC controllers is 60\% lower than that of a centralized MPC (last row of Table \ref{tab:Simulation results}). Using the proposed scheme for the traffic network, not all the agents need to be connected; therefore the communication effort is significantly lower compared to the centralized scheme. CasADi optimization tool is also perfectly appropriate for the distributed plant with less complication and computation. The critical feature of the proposed DMPC algorithm is that it considers the traffic demand and disturbance predictions based on the history of traffic flow, and it distributes the traffic congestion in the links evenly (noticeable in the state graphs of 17 links). \\
\section{CONCLUSIONS}
In this paper, distributed model predictive control strategy was implemented on a nonlinear urban traffic network. The UTN consists of 8 intersections and 17 links. The objective was to reduce the total time spent, the number of vehicles traveling, and the queue length in each road/link. \\ \indent
In this approach, the algorithm considered disturbance and traffic inflow demand predictions. A STARIMA prediction method was used to attain the predictions. The constrained optimization problem was solved using the CasADi tool. \\ \indent
The maximum computation overhead of controllers in the proposed distributed strategy was 60\% lower compared to the classical centralized MPC. Besides, the TTS and the maximum queue length were reduced by 14.3\% and 15.1\%, respectively, and the state and input constraints were totally satisfied throughout the simulation time. For the future, the UTN response to some unpredicted failures, such as actuator failures, can be surveyed practically; e.g., the physical intelligent transportation system testbed of [18].



\begin{thebibliography} {99}
    \bibitem{c1} Yan Y. and Xu C., ``A Development Analysis of China's Intelligent Transportation System," 2013 IEEE International Conference on Green Computing and Communications (GreenCom), pp. 1072-1076, 2013.
    \bibitem{c2} Lin S., B. De Schutter, Y. Xi, Hellendoorn H., ``Efficient network-wide model-based predictive control for urban traffic networks," Transp. Res. Pt. C-Emerg. Technol., Vol. 24, pp. 122–140, 2012. 
    \bibitem{c3} Lin S., B. De Schutter, Xi Y., Hellendoorn H., ``Fast model predictive controllers for large urban road networks via MILP," IEEE Trans. Intell. Transp. Syst., Vol. 12, No. 3, pp. 846–856, 2011.
    \bibitem{c4} Van den Berg M, Hegyi A., De Schutter B., Hellendoorn J., ``Integrated traffic control for mixed urban and freeway networks: A model predictive approach," Eur. J. Transp. Infras. Res., Vol. 7, No. 3, pp. 223-250, 2012.
    \bibitem{c5} Hsieh P., Chen Y., Wu W. and Hsiung P. ``Timing Optimization and Control for Smart Traffic," 2014 IEEE Green Computing and Communications (GreenCom), pp. 9-16, 2014.
    \bibitem{c6} Hardy J. and Liu L., ``Available Forward Road Capacity Detection Algorithms to Reduce Urban Traffic Congestion," 2017 IEEE Green Computing and Communications (GreenCom), pp. 110-119, 2017.  
    \bibitem{c7} Creemers F., Medina A., Lefeber E., Wouw N., ``Design of a supervisory controller for Cooperative Intersection Control using Model Predictive Control," IFAC-PapersOnLine, Vol. 51, Issue 33, pp. 74-79, 2018. 
    \bibitem{c8} Piacentini G., Goatin P., Ferrara A., ``Traffic control via moving bottleneck of coordinated vehicles," IFAC-PapersOnLine, Vol. 51, Issue 9, pp. 13-18, 2018.
    \bibitem{c9} Camponogara E., Santos da Silva R. and de Aguiar M. A. S., ``A distributed dual algorithm for distributed MPC with application to urban traffic control," 2017 IEEE Conference on Control Technology and Applications (CCTA), pp. 1704-1709, 2017.
    \bibitem{c10} Tettamanti T. and Varga I., ``Distributed traffic control system based on model predictive control," Period. Polytech.‐Civ. Eng., Vol. 54, No. 1, pp. 3–9, 2010. 
    \bibitem{c11} De Oliveira L. and Camponogara E., ``Multi‐agent model predictive control of signaling split in urban traffic networks," Transp. Res. Pt. C‐Emerg. Technol., Vol. 18, No. 1, pp. 120–139, 2010.
    \bibitem{c12} Zhou Z., De Schutter B., Lin S., Xi Y., ``Two-Level Hierarchical Model-Based Predictive Control for Large-Scale Urban Traffic Networks," IEEE Transactions on Control Systems Technology, vol. 25, no. 2, pp. 496- 508, 2017. 
    \bibitem{c13} Hu H., Pu Y., hen M., Tomlin C. J., ``Plug and Play Distributed Model Predictive Control for Heavy Duty Vehicle Platooning and Interaction with Passenger Vehicles," 2018 IEEE Conference on Decision and Control (CDC), Miami Beach, FL, USA, 2018.
    \bibitem{c14} Eini R. and Abdelwahed S., ``Distributed Model Predictive Control Based on Goal Coordination for Multi-Zone Building Temperature," 2019 IEEE Green Technologies Conference (GreenTech), Lafayette, LA, 2019.
    \bibitem{c15} Li S., Zheng Y. Distributed model predictive control for plant-wide systems. Wiley; 2015.
    \bibitem{c16} Lin S., Huang H., Zhu D., Wang T., ``The application of space-time ARIMA model on traffic flow forecasting," 2009 International Conference on Machine Learning and Cybernetics,  pp. 3408-3412, 2009.
    \bibitem{c17} Eini R., Linkous L., Zohrabi N., Abdelwahed S., ``A Testbed for a Smart Building: Design and Implementation," Proceedings of the 4th International Workshop on Science of Smart City Operations and Platforms Engineering, April 15-18, Montreal, QC, Canada, 2019. \\
    DOI: 10.1145/3313237.3313296 
    \bibitem{c18} Morrissett A., Eini R., Zaman M., Zohrabi N., Abdelwahed S., ``A Physical Testbed for Intelligent Transportation Systems," Proceedings of the 12th IEEE International Conference on Human System Interaction IEEE HSI 2019, June 25-27, Richmond, VA, USA, 2019. 
    \bibitem{c19} Singh M. G. and Titli A. Systems-decomposition, optimization and control. Berlin: Pergamon Press; 1978.


\end{thebibliography}
\end{document}